\documentstyle[11pt,newpasp,twoside]{article}
\index{fallback accretion disks}
\index{magnetars}
\index{soft gamma-ray repeaters}
\index{anomalous x-ray pulsars}

\def\edcomment#1{\iffalse\marginpar{\raggedright\sl#1\/}\else\relax\fi}
\reversemarginpar

\begin{document}
\title{Propeller vs. Magnetar Concepts for SGR/AXPs}
 \author{Richard E. Rothschild, Richard E. Lingenfelter \& David Marsden}
\affil{Center for Astrophysics and Space Sciences, University of California San Diego
9500 Gilman Dr., La Jolla, CA 92093-0424}

\begin{abstract}
Two lines of thought exist as to the nature of Soft Gamma-ray Repeaters
(SGRs) and Anomalous X-ray Pulsars (AXPs). On the one hand, Duncan \& Thompson (1992)
and Thompson \& Duncan (1995)
propose neutron stars with super-critical
($>10^{14}$ G) magnetic fields, which spin-down the stars and power
the gamma-ray bursts. On the other hand, several authors (van Paradijs, Taam \&
van den Heuvel 1995; Chatterjee, Hernquist \& Narayan 2000; Alpar
2001; Marsden et al. 2001; Menou, Perna \& Hernquist 2001)
propose neutron stars with typical pulsar
magnetic fields ($\sim10^{12}$ G), which are spun-down by magnetospheric
``propeller'' torques from fallback or fossil disks in addition to magnetic
dipole radiation. We discuss these two concepts in light of various
observations.
\end{abstract}

\section{Magnetar \& Fallback Accretion Disk Concepts}

Magnetars, defined to be neutron stars possessing dipole magnetic fields
in excess of the quantum critical value of 4.4$\times$10$^{13}$ Gauss,
constitute a proposed class distinct from radio and x-ray pulsars,
in which magnetic energy, rather than rotational energy, plays the
dominant role in powering emissions. The strong magnetic dipole radiation
(MDR) would spin-down magnetars quite rapidly leaving them with spin periods
of a few seconds after $\sim$10$^3$ years. Repetitive soft gamma-ray
bursts are interpreted as due to crust cracking events in the neutron
star surface, whereas the super busts seen from SGR0525-66 and SGR1900+14
would result from sudden large-scale magnetic reconnection.
Problems replicating the estimated ages of SGR/AXPs in this model have
led to modeling extra sources of torque on the system, but with the
magnetic energy remaining as the dominant power source (Kouveliotou et al. 1999).

Alternatively, the rapid spindown rates, young ages inferred from the SNR ages,
long spin periods clustered around 5-10 s, and $\sim$ 10$^{35}$ erg/s
x-ray luminosities for SGR/AXPs can all be explained by models involving
the propeller effect on inflowing material as the
dominant spindown torque. This material comes from a small accretion
disk formed around the neutron star very early in its life. Such
a disk can form in several ways: from the inner most ejecta material
falling back within a few hours of the initial supernova explosion
(Michel 1988; Chatterjee et al 2000); from the reversal of slower-moving inner ejecta
by the Sedov phase reverse shock relatively soon after the blast wave
hits the progenitor winds (Truelove \& McKee 1999); or from high velocity neutron
stars capturing comoving ejecta (van Paradijs et al. 1995). Only a very small fraction
of the ejecta is needed to form a fossil disk of 10$^{-6} M_\odot$
which is all that is required to explain the spindown of SGR/AXPs via
the propeller mechanism. In this model, the exceedingly rapid spindown
causes crust cracking and subduction to provide both the energy and mechanism for the
very energetic bursts.

\section{Observational Constraints}

Observations of SGRs and AXPs have revealed many of the characteristics
of these objects that must be explained by a successful theory
(Rothschild, Marsden \& Lingenfelter 2002).
Additionally, the theory must predict or be consistent with ideas
of the histories of such objects, such as conditions surrounding
their birth and their galactic inventory. Table~\ref{tab:const}
gives a list of constraints and whether or not they are explained
by either of the two concepts for SGR/AXPs. We discuss each of
these constraints below.

\begin{table}
\caption{Observational Constraints}
\label{tab:const}
\begin{tabular}{llll}
\hline \hline
Constraint & Value & Magnetars & Disks\\
\hline
Spin Period Distribution & Clustered around 5-10 s & No & Yes\\
Spin-Down Rates \& Ages & P/2$\dot{P} \not=$ SNR Age & No & Yes\\
Braking Indices & $\not=$3 & No & Yes\\
Spin-Down Noise & Larger than in Pulsars & ?? & Yes\\
Located in Dense ISM & Opposite than for Pulsars & No & Yes\\
Visibility of Accretion Disk & Very small if at all & Yes & ??\\
Number of Objects & $\sim$10 & ?? & Yes\\
Normal Burst Energy & $\sim10^{41}$ergs & Yes & Yes\\
Super Burst Energy & $\sim10^{44}$ergs & Yes & Yes\\
Burst Durations & $\sim0.2$s & Yes & Yes\\
Abrupt Changes in $\dot{P}$ & $\Delta \dot{P}/P \approx$1 & No & Yes\\
Change in Pulse Profile & Simplified at Superburst & Yes & ??\\
\hline
\end{tabular}
\end{table}

A basic property of AXP/SGRs is their narrow range of spin periods
from 5 to 10 s. Such a clustering is a natural result of the
equilibrium period reached by the propeller effect in low luminosity
accretion disks around neutron stars with the pulsar
distribution of magnetic fields, but
is not consistent with the magnetars, which should show a much
wider range of values, even with field decay.

Another basic property of the AXP/SGRs is their measured spin-down
rates and periods, which give MDR spin-down ages ($P/2\dot{P}$)
expected in the magnetar model that are much
shorter and not consistent with the ages of the associated
supernova remnants. Therefore, another source of torque on
the neutron star must be present in the magnetar model.
Addition of propeller driven spin-down can give ages that
are quite consistent with the associated supernova remnants.
The original magnetar model has been modified to
include a torque component from a relativistic wind in order
to correct age predictions. But such a wind must have nearly
a 100\% efficiency for x-ray production, to be consistent with
the quiescent flux and still require super-critical magnetic fields.
Rothschild, Marsden \& Lingenfelter (2000) have shown that an x-ray production efficiency of
a few percent or less for the wind implies sub-critical fields
consistent with typical pulsar values. The measured $P$s and $\dot{P}$s
of the SGRs together with their ages also give
braking indices which are significantly different than the value
of 3, predicted for MDR alone, but are quit consistent
with that expected from propeller driven spin-down.

The timing noise in AXP/SGRs is much larger than that found in
radio pulsars from MDR, although additional
mechanisms have been proposed by Thompson et al. (2000) that might
account for such noise for magnetars. The timing noise is comparable to that seen in accreting
binary x-ray pulsars (Woods et al. 2000), as would also be expected from
variable accretion in the propeller model.

Most, if not all, of the AXP/SGRs are associated with known supernova
remnants to a high degree of statistical significance (Marsden et al.
2001; see Gaensler et al. 2001 for a contrary opinion). One can use these
SNRs to probe the density of the environment in which the AXP/SGRs
were born. While 80-90\% of neutron star-producing core-collapse SNae
occur in the hot tenuous medium of superbubbles, the SNae associated
with AXPs and SGRs show the opposite tendency, i.e., $>$80\% occur
in the denser ISM. Such higher densities will confine the massive
progenitor winds much closer to the star and these will decelerate the
blast wave much more rapidly and initiate the reverse shock in
the remnant (Truelove \& McKee 1999). This can create the fallback
disks to spin-down the neutron star to the narrow, 5-10 s period range.
Thus, the fallback disk accretion model naturally explains high ISM
density at the birth sites. Nothing in the magnetar model requires, or explains
why the ambient density need be any different than that for neutron stars
in general.

The dense ISM accretion models predict
that a dozen AXP/SGRs have been formed in the last 20 kyrs,
assuming 20\% as the fraction of new neutron stars born in dense
ISM, 10\% for the fraction of massive, rapidly evolving progenitors that
experience mass loss sufficient to form a pushback disk in dense
ISM environments (Marsden et al. 2001), and a SNae rate of 1/40 yr$^{-1}$
over the last 20 kyr. Thus, the
fallback disk scenario can successfully predict the numbers seen.
The magnetar model provides no such estimate.

High sensitivity optical observations of particular SGR/AXPs have
set strong limits on the size of accretion disks assuming the
standard disk model (e.g. Kaplan et al. 2001; Hulleman et al. 2000).
However, Menou et al. (2001) have modeled the dusty, metal-rich disks expected from
supernova fallback and they find that these upper limits are consistent
with such disks. Observations in the infrared are required to test
for the presence of such disks that will cool by very different means
than the standard hydrogen/helium alpha disks.

The total energy of SGR bursts amount to about 10$^{44}$ ergs
for the rare super bursts and about 10$^{41}$
ergs for the weaker more frequent bursts. In the propeller model,
it is proposed that the rapid spin down of the star
creates dynamical stresses within the crust to produce frequent crustal quakes
that provide the energy for the weaker bursts, while rarer, much stronger
quakes from compressive phase changes in subducted crust can provide
the energy for the super bursts. Vibrations excited by these quakes
will be transmitted into magnetospheric Alv\`{e}n waves which accelerate
particles, producing the x-ray/gamma-ray emission. In the magnetar model,
it is proposed that the weaker frequent bursts are also caused by
quakes, but ones resulting from crustal cracking produced by magnetar fields. The
rarer super bursts come from magnetic reconnection.

The short durations, typically about 0.2 s, of most of the weaker
bursts, as well as the impulsive phase of the super bursts, have been
attributed to either the gravitational radiation damping time of neutron star vibrations
(Ramaty, Bussard \& Lingenfelter 1980) which provide an extended energy source in sub-critical
field models, or the storage time for energy in the neutron
star crust (Blaes et al. 1989) for a much briefer energy input in the
magnetar model. Assuming nearly instantaneous injection
of all of the super burst energy into a pair plasma in the magnetar
model, a superstrong magnetic field would be required
in order to contain that energy. In subcritical field models,
where the both the spectral hardness and luminosity of super bursts
can be explained (Ramaty et al. 1981; Lindblom \& Detweiler 1983) by synchrotron emission in
$\sim 10^{12}$ G fields, most of the energy is stored in the neutron
star vibrational modes so that the energy in the radiating pair plasma
can easily be confined by the $\sim10^{12}$ G field.

The spin-down rate of SGR 1900+14 measured
after the super burst on August 27, 2000, increased by a factor
of 2 over that prior to the burst. Such an increase would imply
an increase in magnetic field energy in a magnetar,
which is just the opposite of what would be
expected if the burst were powered by magnetic reconnection
which should reduce the field. For the propeller model, assuming a
``standard" hydrogen disk, Thompson et al. (2000) suggested that the
burst would have disrupted the inner portions of an accretion
disk, reducing spin-down. But a thin, high metallicity fallback
disk could easily survive the burst, because the total energy
deposited in the disk would be much less than its gravitational
binding energy, and heating of the inner edge of the disk can
temporarily increase the propeller torques, as is observed.

\section{Conclusions}

The success of the accretion
models is that they require only the well-studied properties of neutron
stars and supernovae, and they can be applied beyond AXP/SGRs to
clarify contradictions in interpretations of other neutron stars.
These models predict the non-bursting attributes ---
luminosity, spin period, spin-down rate --- as well as the low
number seen in the Galaxy. Spin-down driven quakes can also power both the
repetitive bursting and the super bursts, and the durations of
these bursts are consistent with postquake vibrational damping times.
Direct observations of the disks, however, are needed to establish their existence.

The magnetar model with relativistic winds can also explain both the persistent
and bursting x- and gamma-ray emission
from SGRs, and the spin-down of both the SGRs and AXPs, if the wind x-ray emission
efficiency is near 100\%.  The magnetar model, however, does not
explain the clustering of spin periods observed in these sources, even with magnetic
field decay. Theoretical arguments suggest that magnetic fields
can exist far above the critical field, but observational evidence
from all of the radio pulsars, whose implied fields from $P$ and $\dot{P}$
span over 5 orders of magnitude, show a clear cutoff just short of
the critical field.

\end{document}